\documentclass[showpacs]{revtex4}
\usepackage{graphicx}

\begin{document}

\title{Mean Field Effect on $J/\psi$ Production in Heavy Ion Collisions}
\author{Baoyi Chen, Kai Zhou, and Pengfei Zhuang}
\affiliation{Physics Department, Tsinghua University, Beijing
100084, China}
\date{\today}

\begin{abstract}
The mass shift effect at finite temperature on $J/\psi$ production
in relativistic heavy ion collisions is calculated in a detailed
transport approach, including both mean field and collision terms.
While the momentum-integrated nuclear modification factor $R_{AA}$
is not sensitive to the mass shift, the reduced threshold for
$J/\psi$ regeneration in the quark-gluon plasma leads to a
remarkable enhancement for the differential $R_{AA}$ at low
transverse momentum.
\end{abstract}
\pacs{25.75.-q, 12.38.Mh, 24.85.+p}
\maketitle

$J/\psi$ is a tightly bound state of charm quarks $c$ and $\overline
c$, its dissociation temperature $T_d$ in hot medium is much higher
than the critical temperature $T_c$ of the deconfinement phase
transition. Therefore, the measured $J/\psi$s in nuclear collisions
at Relativistic Heavy Ion Collider (RHIC) and Large Hadron Collider
(LHC) carry the information of the hot medium and has long been
considered as a signature~\cite{matsui} of the new state of matter,
the so-called quark-gluon plasma (QGP)~\cite{qgp}.

The $J/\psi$ properties are significantly affected by the
deconfinement phase transition. From the calculations with quantum
chromodynamics (QCD) sum rules~\cite{megias,morita} and QCD
second-order Stark effect~\cite{lee}, both the $J/\psi$ width and
mass are largely changed in a static hot medium. For a dynamically
evolutive QGP phase created in the early stage of heavy ion
collisions, the width is induced by $J/\psi$ decay like the gluon
dissociation $g+J/\psi \to c +\overline
c$~\cite{na50,blaizot,capella,hufner,polleri,bratkovskaya,zhuang,zhu,wong},
and the mass shift comes from the mean field effect of the medium.
Since the mass shift for a heavy quark system is expected to be weak
at low temperature, it is neglected in almost all the model
calculations. However, in the region above and close to the critical
temperature, there is a sudden change in the mass of $J/\psi$. For
instance, at temperature $T/T_c=1.1$ the mass shift $\delta
m_{J/\psi}=m_{J/\psi}(T)-m_{J/\psi}(0)$ can reach $-(100-200)$
MeV~\cite{megias,morita,lee}, which is already comparable with the
mass change for light hadrons~\cite{leupold} and should have
remarkable consequence in $J/\psi$ production. In this paper, we
study the hot nuclear matter effect on $J/\psi$ production in heavy
ion collisions at RHIC and LHC energies, including not only the
gluon dissociation but also the mean field.

Let's first qualitatively estimate the mean field effect on the
$J/\psi$ distribution. The decreased mass reduces the threshold
value for the $J/\psi$ production in QGP, and should result in an
enhancement for the $J/\psi$ yield. Secondly, the attractive force
between the inhomogeneous medium and $J/\psi$,
\begin{equation}
\label{force}
{\bf F}({\bf x},{\bf p}) = -{m_{J/\psi}\over
E_{J/\psi}}{\bf \nabla}m_{J/\psi}=-{m_{J/\psi}\over
E_{J/\psi}}{\partial m_{J/\psi}\over \partial T}{\bf \nabla}T
\end{equation}
with $J/\psi$ energy $E_{J/\psi}=\sqrt{m_{J/\psi}^2+{\bf p}^2}$,
pulls $J/\psi$ to the center of the fireball and kicks $J/\psi$ to a
lower momentum region. This will lead to an enhancement at low
momentum and a reduction at high momentum.

\begin{figure}[!hbt]
\includegraphics[width=0.4\textwidth]{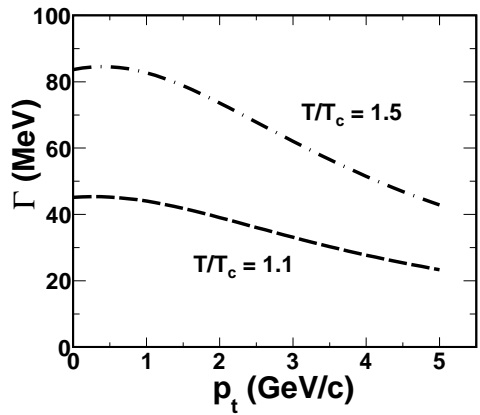}
\caption{The $J/\psi$ decay width as a function of transverse
momentum $p_t$ at temperature $T/T_c = 1.1$ and $1.5$. The medium
velocity is fixed as $v_{QGP} = 0.5$ and its direction is chosen as
the same as the $J/\psi$ momentum. }
\label{fig1}
\end{figure}

Since charmonia are so heavy and difficult to be thermalized in hot
medium, we use a detailed transport approach~\cite{yan} to describe
their distribution functions $f_\Psi({\bf p},{\bf x},\tau|{\bf b})$
in the phase space at fixed impact parameter ${\bf b}$ of a
nucleus-nucleus collision,
\begin{equation}
\label{transport}
{\partial f_\Psi\over \partial \tau}+{{\bf p}\over
E_\Psi}\cdot{\bf \nabla}_x f_\Psi+{\bf F}\cdot{\bf \nabla}_p f_\Psi
=-\alpha_\Psi f_\Psi+\beta_\Psi,
\end{equation}
where ${\bf \nabla}_x$ and ${\bf \nabla}_p$ indicate three
dimensional derivatives in coordinate and momentum spaces.
Considering the fact that in proton-proton collisions the
contribution from the decay of the excited charmonium states to the
$J/\psi$ yield is about $40\%$~\cite{herab}, we must take transport
equations not only for the ground state $\Psi=J/\psi$ but also the
excited states $\Psi=\psi'$ and $\chi_c$. The collision terms on the
right hand side are from the charmonium inelastic interaction with
the medium. The lose term $\alpha$~\cite{zhu,yan} arising from the
gluon dissociation process controls the $J/\psi$ suppression, and
the gain term $\beta$~\cite{yan} related to $\alpha$ by detailed
balance governs the $J/\psi$
regeneration~\cite{pbm,gorenstein,thews1,rapp,thews2,zhao} in the
QGP phase. The gluons and charm quarks in $\alpha$ and $\beta$ are
assumed to be thermalized at RHIC and LHC energies. The cross
section for the gluon dissociation $\sigma_\Psi(0)$ in vacuum has
been calculated through the method of operator production
expansion~\cite{peskin,bhanot}, and its value at finite temperature
can be obtained from its classical relation to the charmonium size
$r_\Psi$, $\sigma_\Psi(T)=\langle r_\Psi^2(T)\rangle/\langle
r_\Psi^2(0)\rangle\sigma_\Psi(0)$, where the averaged $r$-square can
be derived from the Schr\"odinger equation~\cite{satz} with lattice
simulated potential~\cite{karsch,shuryak}. In Fig.\ref{fig1}, we
show the $J/\psi$ decay rate $\Gamma\equiv\alpha_{J/\psi}$ as a
function of transverse momentum at fixed temperature $T/T_c=1.1$ and
$1.5$. In the calculation here we have chosen a typical medium
velocity $v_{QGP}=0.5$ and assumed that $\vec v_{QGP}$ and $J/\psi$
momentum $\vec p$ have the same direction. In the vicinity of the
phase transition, the width is not sensitive to the momentum, and
the amplitude is in qualitative agreement with the result calculated
from QCD sum rules~\cite{morita}. When the temperature increases
from $1.1T_c$ to $1.5T_c$, the width increases by a factor of about
2.
%%%%%%%%%%%%%%%%%%%%%%%%%%%%%%%%%%%%%%%%%%%%%%%%%%%%%%%%%%%%%%%%%

Different from the previous study, we consider here not only the
inelastic processes, but also a mean field term ${\bf F}\cdot{\bf
\nabla}_p f_\Psi$. The local temperature $T({\bf x},\tau)$ appeared
in the elastic and inelastic terms are obtained from the
hydrodynamic equations. At RHIC and LHC energies, one can take the
$2+1$ dimensional version~\cite{heinz1,hirano1,zhu} of the
relativistic hydrodynamic equations
\begin{equation}
\label{hydro}
\partial_\mu T^{\mu\nu}=0,\ \ \partial_\mu N^\mu=0
\end{equation}
to describe the space-time evolution of the QGP, where $T_{\mu\nu}$
and $N_\mu$ are the energy-momentum tensor and baryon number current
of the system. In our numerical calculation, the QGP formation time
is chosen to be $\tau_0=0.6$ fm/c and the initial thermodynamics is
determined by the colliding energy and the nuclear
geometry~\cite{zhu,hirano2}.

The charm quark mass $m_c$ is determined by the charmonium mass
$m_\Psi$ and its binding energy $\epsilon_\Psi$~\cite{satz},
$m_c=\left(m_\Psi-\epsilon_\Psi\right)/2$. The cold nuclear matter
effects on $J/\psi$ production, such as nuclear
absorption~\cite{gerschel}, Cronin effect~\cite{gavin,hufner2} and
shadowing effect~\cite{vogt} can be reflected in the initial
charmonium distributions $f_\Psi({\bf p},{\bf x},\tau_0|{\bf b})$.
In the following we neglect the nuclear absorption at RHIC and LHC
energies due to the small QGP formation time, and consider the
Cronin effect with a Gaussian smearing treatment~\cite{zhao,liu}.

The transport equation with the mean field force can not be solved
analytically. Supposing that the very small elastic cross section
for charmonia at hadron level is still true at parton level, the
mean field term can be considered as a perturbation, and the
transport equation can be solved perturbatively. Fully neglecting
the mass shift, the zeroth-order transport equation can be strictly
solved with the solution~\cite{yan}
\begin{equation}
\label{f0}
f_\Psi^{(0)}({\bf p},{\bf x},\tau|{\bf b}) = f_\Psi({\bf
p},{\bf x}_0,\tau_0|{\bf b}) e^{-\int_{\tau_0} ^\tau d\tau_1
\alpha_\Psi({\bf p},{\bf x}_1,\tau_1|{\bf b})} +\int_{\tau_0} ^\tau
d\tau_1 \beta_\Psi({\bf p},{\bf x}_1,\tau_1|{\bf
b})e^{-\int_{\tau_1}^\tau d\tau_2\alpha_\Psi({\bf p},{\bf
x}_2,\tau_2|{\bf b})}
\end{equation}
with the coordinate shift ${\bf x}_n={\bf x}-{\bf
p}/E_\Psi(\tau-\tau_n)$ which reflects the leakage
effect~\cite{matsui,hufner3} during the time period $\tau-\tau_n$.
The first term on the right hand side is the contribution from the
initial production, it suffers from the gluon dissociation during
the whole evolution of QGP. The second term arises from the
regeneration and suffers also the suppression from the production
time $\tau_1$ to the end of QGP.

Taking into account the fact that the dissociation temperatures for
the excited states are around $T_c$~\cite{satz}, we do not consider
their mass shifts in QGP. Therefore, for the transport equations for
$\psi'$ and $\chi_c$, their masses are temperature independent and
the mean field forces disappear, the zeroth-order solutions become
the full distributions, $f_{\psi'}=f_{\psi'}^{(0)}$ and
$f_{\chi_c}=f_{\chi_c}^{(0)}$.

We now consider the mean field effect on the $J/\psi$ distribution
$f_{J/\psi}$. We extract the mass shift from the QCD second-order
Stark effect~\cite{lee},
\begin{eqnarray}
\label{mass}
\delta
m_{J/\psi}(T)&=&m_{J/\psi}(T)-m_{J/\psi}(0)\\
&=&-{7\pi^2\over 18}{a^2\over e}\left[{2\over 11}M_0(T)+{3\over
4}{\alpha_s(T)\over \pi}M_2(T)\right]\nonumber,
\end{eqnarray}
which is used as input for the spectral function analysis with QCD
sum rules~\cite{morita}, where the constants are taken as $a=0.271$
fm and $e=640$ MeV, the coupling constant $\alpha_s$ is temperature
dependent~\cite{lee,morita}, and the functions $M_0=\epsilon-3P$ and
$M_2=\epsilon+P$ determined by the energy density $\epsilon$ and
pressure $P$ of the medium are extracted from the lattice QCD
simulations~\cite{boyd}. Leaving out first the mean field force
induced by the inhomogeneous property of the fireball but keeping
the temperature dependence of the mass, the solution of the
transport equation has the same form with the zeroth-order
distribution, the only difference is the replacement of
$m_{J/\psi}(0)\rightarrow m_{J/\psi}(T)$,
\begin{equation}
\label{f1}
f_{J/\psi}^{(1)}=f_{J/\psi}^{(0)}\big|_{m_{J/\psi}(0)\rightarrow
m_{J/\psi}(T)}.
\end{equation}

If the effect of the mean field force is not very strong, we may
infer that it is not such a bad approximation to replace
$f_{J/\psi}$ in the mean field term by the known $f_{J/\psi}^{(1)}$,
the approximate transport equation
\begin{equation}
\label{transport1}
{\partial f_{J/\psi}\over \partial \tau}+{{\bf
p}\over E_{J/\psi}}\cdot{\bf \nabla}_x f_{J/\psi} =-\alpha_{J/\psi}
f_{J/\psi}+\beta_{J/\psi}-{\bf F}\cdot{\bf \nabla}_p
f_{J/\psi}^{(1)}
\end{equation}
is then similar to the equation for $f_{J/\psi}^{(1)}$ but with a
replacement for the regeneration
\begin{equation}
\label{regeneration}
\beta_{J/\psi}\to \beta_{J/\psi}-{\bf
F}\cdot{\bf \nabla}_p f_{J/\psi}^{(1)}.
\end{equation}
This means that the mean field force can be considered as an
effective regeneration, which is not from the coalescence of heavy
quarks but due to the $J/\psi$ mass shift. The approximate solution
so obtained is known as the second-order $J/\psi$ distribution,
\begin{equation}
\label{f2}
f_{J/\psi}^{(2)}({\bf p},{\bf x}, \tau|{\bf b}) =
f_{J/\psi}^{(1)}({\bf p},{\bf x},\tau|{\bf b})-\int_{\tau_0}^\tau d
\tau_1 \left[{\bf F}({\bf x}_1)\cdot {\bf \nabla}_p
f_{J/\psi}^{(1)}({\bf p},{\bf x}_1,\tau_1|{\bf b})\right]
e^{-\int_{\tau_1}^\tau d\tau_2\alpha({\bf p},{\bf x}_2,\tau_2|{\bf
b})}.
\end{equation}

Substituting the obtained $f_{J/\psi}^{(2)}$ into the mean field
term, and solving the transport equation again, we can derive the
third-order distribution function $f_{J/\psi}^{(3)}$. With the
similar way, the distribution to the $n$-th order can be generally
expressed as a series of the mean field force,
\begin{eqnarray}
\label{fn} f_\Psi^{(n)}({\bf p},{\bf x}, \tau|{\bf b}) &=&
f_\Psi^{(1)}({\bf p},{\bf x},\tau|{\bf
b})-\sum_{m=1}^{n-1}(-1)^{m-1}\int_{\tau_0} ^\tau
d\tau_1\int_{\tau_0} ^{\tau_1}d\tau_2
\cdots\int_{\tau_0}^{\tau_{m-1}}d\tau_m\nonumber\\
&\times& \left[\left(\prod_{i=1}^m{\bf F}({\bf x}_i)\cdot{\bf
\nabla}_p\right) f_\Psi^{(1)}({\bf p},{\bf x}_m,\tau_m|{\bf
b})\right]e^{-\int_{\tau_m}^\tau d\tau^{\prime}\alpha({\bf p},{\bf
x}^{\prime},\tau^{\prime}|{\bf b})}.
\end{eqnarray}
Note again that the mass shift affects the $J/\psi$ production in
two aspects. The reduced threshold, which is considered already in
the first-order distribution $f_{J/\psi}^{(1)}$, makes the
production more easily, and the attractive force between $J/\psi$
and the matter, which is included only in the higher-order
distributions $f_{J/\psi}^{(n)}\ (n=2,3,\cdots)$, makes a momentum
shift and leads to a low $p_t$ enhancement and a corresponding high
$p_t$ suppression.

\begin{figure}[!hbt]
\includegraphics[width=0.4\textwidth]{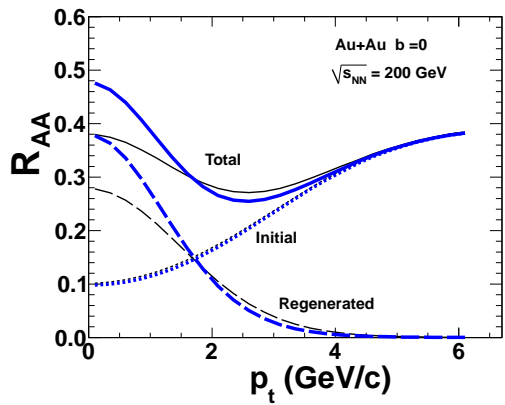}
\caption{(color online) The differential nuclear modification factor
$R_{AA}(p_t)$ as a function of transverse momentum $p_t$ in Au+Au
collisions at impact parameter $b=0$ and top RHIC energy $\sqrt
{s_{NN}}=200$ GeV. The initial production, the regeneration, and the
total are shown by dotted, dashed and solid lines. The thick and
thin lines are the calculations with and without considering the
mean field effect.}\label{fig2}
\end{figure}

By integrating the charmonium distribution function $f_\Psi({\bf
p},{\bf x},\tau|{\bf b})$ over the hyper surface of hadronization
determined by $T({\bf x},\tau)=T_c$, and considering the decay of
the excited states to the ground state, we can calculate the number
$N_{AA}$ of survived $J/\psi$s in a heavy ion collision. Let's first
examine the differential nuclear modification factor
$R_{AA}(p_t)=N_{AA}(p_t)/\left(N_c N_{pp}(p_t)\right)$ as a function
of transverse momentum $p_t$ in mid rapidity region, where $N_{pp}$
is the number of $J/\psi$s in a nucleon-nucleon collision and $N_c$
the number of nucleon-nucleon collisions. To see the largest mean
field effect, we calculate first $R_{AA}(p_t)$ in central collisions
with ${\bf b}=0$ where the formed fireball is hot and large and the
surviving time is long. Fig.\ref{fig2} shows our numerical results
for Au+Au collisions at top RHIC energy $\sqrt {s_{NN}} = 200$ GeV.
The thick and thin lines indicate our results with and without
considering the $J/\psi$ mass shift, by taking $f_{J/\psi}$ and
$f_{J/\psi}^{(0)}$, respectively. The dotted, dashed and solid lines
are the calculations with initial production only, regeneration only
and the total.

From our numerical results, the series $f_{J/\psi}^{(n)}\
(n=0,1,2,\cdots)$ converges rapidly, and the deviation
$\left|R_{AA}^{(2)}-R_{AA}^{(1)}\right|/\left|R_{AA}^{(1)}\right|<3\%$
is valid in any case. This confirms our qualitative estimation that
the mean field effect is mainly reflected in the change in the
threshold, the attractive force is a second order effect and the
higher order corrections with $n>2$ can be safely neglected. For the
following numerical calculations we will take
$f_{J/\psi}=f_{J/\psi}^{(2)}$. Since the initial production has
ceased before the formation of the hot medium, the change in the
threshold does not affect it, and the correction is only from the
small mean field force. That is the reason why the initial
contribution shown in Fig.\ref{fig2} is almost independent of the
mass shift. However, the regeneration happens in the hot medium, it
is affected by both the reduced threshold and the mean field force,
the overall correction should be much larger in comparison with the
correction to the initial production. Considering the fact that the
regenerated $J/\psi$s in the early stage of the QGP will be eaten up
by the hot medium and only the $J/\psi$s regenerated in the later
stage of the QGP can survive, the enhanced regeneration is mainly in
the low $p_t$ region, as shown in Fig.\ref{fig2}. At $p_t=0$, the
total $R_{AA}$ goes up from $0.38$ to $0.48$, the enhancement is
$26\%$. Since some $J/\psi$s are kicked to low $p_t$ region by the
mean field force, there is a little reduction of $R_{AA}$ in the mid
$p_t (2-3$ GeV) region. At high enough $p_t$, there is almost no
effect of the mass shift, and the $J/\psi$ distribution is dominated
by the perturbative QCD in the initial production. In comparison
with our previous calculations~\cite{liu2,zhou}, the Gaussian
smearing treatment~\cite{zhao,liu} for the Cronin effect used here
reduces the $R_{AA}$ at high $p_t$.
%%%%%%%%%%%%%%%%%%%%%%%%%%%%%%%%%%%%%%%%%%%%%%%%%%%%%%%%%%%%%%%%%
%%%%%%%%%%%%%%%%%%%%%%%%%%%%%%%%%%%%%%%%%%%%%%%%%%%%%%%%%%%%%%%%%
\begin{figure}[!hbt]
\includegraphics[width=0.4\textwidth]{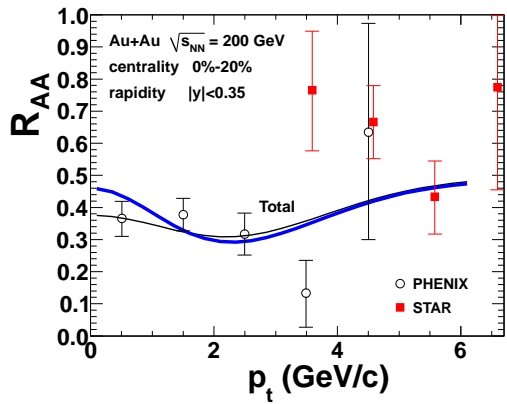}
\caption{(color online) The differential nuclear modification factor
$R_{AA}(p_t)$ as a function of transverse momentum $p_t$ in Au+Au
collisions at impact parameter $b=4.2$ fm and top RHIC energy $\sqrt
{s_{NN}}=200$ GeV. The data are from the PHENIX~\cite{phenix} and
STAR~\cite{star} collaborations, and the thick and thin lines are
the full calculations with and without considering the mean field
effect.}\label{fig3}
\end{figure}

To compare with the experimental data, we show in Fig.\ref{fig3} the
RHIC data for Au+Au collisions at centrality
$0\%-20\%$~\cite{phenix,star} and our calculation at $b=4.2$ fm.
With decreasing centrality, the fireball temperature drops down and
the mean field effect should be gradually reduced. However, from
$b=0$ to $4.2$ fm, the change in the mean field is small, and the
model calculations are almost the same. In our calculation we used
the EKS98 parton distribution function~\cite{eks98} for the
shadowing evolution and incorporated it with our transport model
through the initial distribution. We show in Fig.\ref{fig3} only the
total calculation, the results with and without the mean field
effect can both describe the data reasonably well. Since the current
data are still with large uncertainty even in low $p_t$ region, we
need more precise data to distinguish the mean field effect in the
distribution.

\begin{figure}[!hbt]
\includegraphics[width=0.4\textwidth]{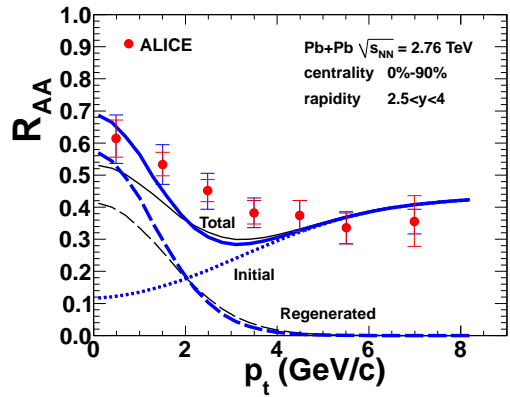}
\caption{(color online) The differential nuclear modification factor
$R_{AA}(p_t)$ as a function of transverse momentum $p_t$ in Pb+Pb
collisions at LHC energy $\sqrt {s_{NN}}=2.76$ TeV. The data are
from the ALICE collaboration~\cite{alice} at centrality $0\%-90\%$
and at forward rapidity $2.5<y<4$, and the model calculation is for
the impact parameter $b=7.2$ fm. The initial production, the
regeneration, and the total are shown by dotted, dashed and solid
lines, and the thick and thin lines are the calculations with and
without considering the mean field effect. }\label{fig4}
\end{figure}

Note that in the above calculations the shadowing effect reduces the
charm quark number and in turn the $J/\psi$ regeneration in the
medium. Since the parton momentum fraction $x_F$ at RHIC energy does
not lie in the remarkable shadowing region, the shadowing effect on
$J/\psi$ production is not remarkable. However, $x_F$ at LHC energy
becomes very small and lies in the strong shadowing region. This
would give rise to a considerable reduction of the $J/\psi$
regeneration. We calculated the $J/\psi$ $R_{AA}(p_t)$ in Pb+Pb
collisions at LHC energy $\sqrt {s_{NN}}=2.76$ TeV and compared it
with the data from the ALICE collaboration~\cite{alice}, as shown in
Fig.\ref{fig4}. Besides the shadowing effect, another important
effect one should include in the comparison with the ALICE data is
the B meson decay, which leads to the non-prompt part in the
inclusive $J/\psi$ yield. We use the experimentally measured p+p
data from the CDF~\cite{cdf}, CMS~\cite{cms} and ATLAS~\cite{atlas}
collaborations to estimate the decay contribution and take the B
meson quench in the medium. The contribution from the B meson decay
is important at high $p_t$ but the influence is small at low $p_t$.
The charm quark production cross section $\sigma_{c\bar c}=0.38$ mb
comes from the combination of the FONLL calculation~\cite{fonll} and
the p+p data~\cite{pp}. From the numerical calculation, the
shadowing effect at LHC results in a $25\%$ reduction of the
$J/\psi$ regeneration, and this makes a better agreement between the
calculation including shadowing effect and the data at low $p_t$,
see Fig.\ref{fig4}.
%%%%%%%%%%%%%%%%%%%%%%%%%%%%%%%%%%%%%%%%%%%%%%%%%%%%%%%%%%%%%%%%%

While the mean field effect changes significantly the differential
nuclear modification factor at low transverse momentum, it does not
affect the global yield remarkably. From Figs.\ref{fig2}-\ref{fig4},
the most important mean field effect is at $p_t=0$ and the effective
region is around $p_t=0.5$ GeV which is much less than the averaged
$J/\psi$ transverse momentum $\langle p_t\rangle\simeq 2-3$ GeV at
RHIC and LHC energies. Therefore, the $p_t$-integrated nuclear
modification factor $R_{AA}(N_p)$ as a function of the number $N_p$
of participant nucleons becomes not sensitive to the mean field
effect. From our numerical calculations, the mass shift induced
enhancement of $R_{AA}(N_p)$ is very small in peripheral and
semi-central collisions and less than $10\%$ even in most central
collisions. Different from $R_{AA}$ which is a summation of the
initial production and regeneration, the averaged transverse
momentum square $\langle p_t^2\rangle$ is governed by the fraction
of the regeneration~\cite{liu2,zhou}. We found that the enhanced
$J/\psi$ yield at low $p_t$ leads to a less than $10\%$ suppression
of $\langle p_t^2\rangle$ in most central collisions.

In summery, we developed a perturbative expansion to study the mean
field effect on $J/\psi$ production in heavy ion collisions. Taking
the mean field term as a perturbation, we analytically solved the
$J/\psi$ transport equation with elastic and inelastic terms to any
order and found a rapid convergence of the perturbative expansion.
The initial $J/\psi$ production, which happens before the formation
of QGP, is not sensitive to the mean field force, but the continuous
regeneration, which happens in the evolution of QGP, is
significantly affected by the mean field. As a result, the
differential nuclear modification factor of $J/\psi$ is enhanced at
low $p_t$ in heavy ion collisions at RHIC and LHC.

\vspace{0.2cm}
\appendix {\bf Acknowledgement:} We thank useful
discussions with Yunpeng Liu, Zhen Qu and Nu Xu. The work is
supported by the NSFC (Grant Nos. 10975084 and 11079024) and RFDP
(Grant No.20100002110080 ).

\end{document}